\documentclass[aps,prl,twocolumn,preprintnumbers,superscriptaddress,dblfloatfix,nofootinbib]{revtex4-1}
\usepackage[utf8]{inputenc}
\usepackage{lmodern}
\usepackage{amsmath,amssymb}
\usepackage{graphicx}
\usepackage{url}
\usepackage{color}
\usepackage{enumitem}
\usepackage{subfigure}
\usepackage[dvipsnames]{xcolor}
\usepackage[colorlinks=true,breaklinks=true]{hyperref}
\hypersetup{allcolors=[rgb]{0.0 0.0 0.6},linkcolor=[rgb]{0.75 0.05 0.05}}
\usepackage{bm}
\usepackage{epsfig}
\usepackage{amsmath}
\usepackage{amssymb}
\usepackage{slashed}
\usepackage{color}
\usepackage{accents}
\usepackage[dvipsnames]{xcolor}
\usepackage[colorlinks=true,breaklinks=true]{hyperref}
\hypersetup{allcolors=[rgb]{0.0 0.0 0.6},linkcolor=[rgb]{0.75 0.05 0.05}}

\renewcommand\[{\left[}

\newcommand{\exclude}[1]{}

\begin{document}
\preprint{IPMU22-0019} 
\preprint{YITP-22-95} 

\title{Gravitational Waves from Rapid Structure Formation on Microscopic Scales before Matter-Radiation Equality}
	
\author{Marcos M.  Flores} 
\affiliation{Department of Physics and Astronomy, University of California, Los Angeles \\ Los Angeles, California, 90095-1547, USA} 
\author{Alexander Kusenko} 
\affiliation{Department of Physics and Astronomy, University of California, Los Angeles \\ Los Angeles, California, 90095-1547, USA}
\affiliation{Kavli Institute for the Physics and Mathematics of the Universe (WPI), UTIAS \\The University of Tokyo, Kashiwa, Chiba 277-8583, Japan}
\affiliation{Theoretical Physics Department, CERN, 1211 Geneva 23, Switzerland}
\author{Misao Sasaki} 
\affiliation{Kavli Institute for the Physics and Mathematics of the Universe (WPI), UTIAS \\The University of Tokyo, Kashiwa, Chiba 277-8583, Japan}
\affiliation{Center for Gravitational Physics and Quantum Information, Yukawa Institute for
Theoretical Physics,\\
Kyoto University, Kyoto 606-8502, Japan}
\affiliation{Leung Center for Cosmology and Particle Astrophysics,
National Taiwan University,\\
Taipei 10617, Taiwan}
	
\date{\today}
	
\begin{abstract}
The existence of scalar fields can be probed by observations of stochastic gravitational waves.  Scalar fields mediate attractive forces, usually  stronger than gravity, on the length scales shorter than their Compton wavelengths, which can be non-negligible in the early Universe, when the horizon size is small.  These attractive forces exhibit an instability similar to the gravitational instability, only stronger.  They can, therefore, lead to the growth of structures in some species.  We identify a gravitational waves signature of such processes and show that it can be detected by future gravitational waves experiments. 

\end{abstract}
\maketitle
	

Most of the elementary particles discovered in the past  century have Compton wavelengths that are much smaller than the size of the atom, and, therefore, they cannot mediate any long-range forces between atoms.  However, in the early Universe with the total energy density $\rho$, any force mediated by particles with masses smaller than $\rho^{1/2}/M_{\rm Pl}$, where $M_{\rm Pl} $ is the reduced Planck mass, acts  as a long-range force on the scales of the Hubble horizon at that time.  Of particular interest are the forces that mediate attractive interactions between particles, such as the Yukawa forces.  The Yukawa interactions are always attractive, and, unless the coupling is unusually small, such forces are much stronger than gravity.  This has a profound effect on the dynamics of any particles participating in such interactions.   The attractive nature of the force implies an instability similar to gravitational instability, and the growth of structures in the distribution of density of some particle species is possible even in the radiation dominated Universe~\cite{Amendola:2017xhl,Savastano:2019zpr, Casas:2016duf,Flores:2020drq, Domenech:2021uyx, Flores:2021jas}.  This growth of isocurvature overdensities can lead to bound states~\cite{Wise:2014jva,Gresham:2017cvl,Gresham:2018rqo} or, accompanied by the radiative cooling caused by the same Yukawa interaction, can lead to the formation of primordial black holes (PBHs)~\cite{Flores:2020drq,Flores:2021jas}.

We will explore a generic observable  manifestation of the structure formation due to the presence of the attractive forces in the early Universe, namely, stochastic gravitational waves (GWs).  We will show that such an event in the early Universe during either a radiation dominated (RD) era or an intermediate matter dominated (IMD) era can produce a signal detectable by the future GW detectors. The possibility of an IMD era has been considered, in particular, in connection with the formation of PBHs~\cite{Khlopov:1980mg, Polnarev1982era, Alabidi:2009bk, Alabidi:2013lya, Alabidi:2013lya}. 

While previous work has demonstrated that gravitational forces alone can generate GWs~\cite{Dalianis:2020gup, Jedamzik:2010hq, Domenech:2021and}, here we will explore the effects of a force significantly stronger than gravity.

The formation of dark matter overdensities due to a long-range scalar force has been examined in many different contexts~\cite{Amendola:2017xhl, Flores:2020drq, Domenech:2021uyx, Savastano:2019zpr}. As in Ref.~\cite{Flores:2020drq}, we consider a heavy fermion $\psi$ interacting with a scalar field $\chi$:
\begin{equation}
\mathcal{L} \supset \frac{1}{2}m_\chi^2\chi^2 - y\chi\bar{\psi}\psi + \cdots
.
\end{equation}
We will assume that the scalar is either massless or very light, $m_\chi\ll m_\psi^2/M_{\rm Pl}$, where $M_{\rm Pl} \approx 2\times 10^{18}$ GeV. Further, we assume that the fermions $\psi$ are either stable or have a total decay width $\Gamma \ll m_\psi^2/M_{\rm Pl}$ which ensures there is a cosmological epoch where the $\psi$ particles can become nonrelativistic, decoupled from equilibrium, and interact via long-range force mediated by the $\chi$ field. In particular, we require that the annihilation cross section $\Gamma_{\psi\psi\to\chi\chi} \ll H$. 

During a RD era, gravity alone only allows for logarithmic growth of density perturbations whereas in an IMD era, gravitational forces cause $\delta(x,t) = \delta\rho/\bar{\rho}$ to grow linearly with scale factor~\cite{Mukhanov:2005sc}, i.e., $\delta\propto a$. In either case, the presence of a long-range ``fifth force" stronger than gravity causes the fluctuations to grow at a much faster rate. We note that the scalar force does not couple to the mass density, but to the number density of $\psi$. The strength of this force can be characterized by its relative strength with gravity, $\xi\equiv y M_{\rm Pl}/m_\psi \gg 1$. This causes fluctuations $\Delta(x,t) = \Delta n_\psi/\bar{n}_\psi$ to grow more rapidly so long as the $\psi$ is decoupled from radiation so that pressure can be neglected.

In Fourier space, the growth of these perturbations is described by the system of coupled equations~\cite{Gradwohl:1992ue, Gubser:2004uh, Nusser:2004qu, Amendola:2017xhl, Domenech:2021uyx},
\begin{align}
\ddot{\delta}_k + 2H\dot{\delta}_k 
-
\frac{3}{2}H^2
(\Omega_{\rm rad}\delta_k + \Omega_\psi\Delta_k) &= 0\\
\ddot{\Delta}_k + 2H\dot{\Delta}_k
-
\frac{3}{2}H^2
\left[
\Omega_{\rm rad}\delta_k + \Omega_\psi(1 + \xi^2)\Delta_k
\right] &= 0,
\label{eq:GrowthEq}
\end{align}
where $\Omega_{\rm rad} = \rho_{\rm rad}/(\rho_{\rm rad} + \rho_\psi)$ and $\Omega_m = \rho_\psi/(\rho_{\rm rad} + \rho_\psi)$ are the radiation and matter components respectively such that $\Omega_{\rm rad} + \Omega_m = 1$ and $H$ is the Hubble parameter $H\equiv \dot{a}/a$. Assuming that radiation and the $\psi$ particles are the only relevant energy components, then the time dependence of the energy fractions are
\begin{equation}
\Omega_{\rm rad} = \frac{1}{1 + (t/t_{\rm eq})^{s}},
\qquad
\Omega_\psi = \frac{1}{1 + (t_{\rm eq}/t)^{s}}
\end{equation}
where $t_{\rm eq}$ is the matter-radiation equality time when the $\psi$ component comes to dominate and $s$ is defined as the power $a\propto t^s$ such that $s = 1/2$ for radiation domination and $s = 2/3$ for matter domination. We can examine the solution of Eq. \eqref{eq:GrowthEq} in the limit that $\xi\gg 1$ in either era
\begin{equation}
\Delta_k \propto a^p,
\quad
p =
\begin{cases}
\sqrt{6(1 + \xi^2)}& {\rm RD}\\[0.25cm]
\frac{1}{4}
\left(
\sqrt{1 + 24(1 + \xi^2)} - 1
\right) & {\rm MD}
\end{cases}
\end{equation}
where $p$ can be much larger than 1. This growth continues until the density contrast $\Delta_k$ enters the nonlinear regime when $\Delta_k\simeq 1$.

Our discussion of the evolution of the overdensities and their subsequent collapse will closely follow the work of Dalianis and Kouvaris~\cite{Dalianis:2020gup}. Rather than utilizing the formalism of induced GWs from a decaying gravitational potential with time, Ref. \cite{Dalianis:2020gup} applies an alternative analytic approach based on the Zel'dovich approximation for the nonlinear evolution of density perturbations~\cite{Zeldovich:pancake}.

As described by Zel'dovich, deviations from spherical symmetry of a halo of pressureless, self-gravitating gas will cause instability, leading to flattening in one particular direction. These objects, often referred to as Zel'dovich pancakes, may lead to the formation of black holes if the final configuration satisfies hoop conjectures~\cite{Thorne:1972ji, Misner:1973prb}. If not, the system will continue to collapse and go through phase crossings and oscillations which lead to a virialized halo. We will be applying the Zel'dovich approximation instead to a system of pressureless gas interacting via a Yukawa interaction. Functionally, a Yukawa interaction is identical to gravity, with both being having $1/r$ potentials. Thus, we expect the Zel'dovich approximation to hold similarly in our scenario.

In the absence of an asymmetry in the $\psi$ component, this leads to the destruction of the halo itself through $\bar{\psi}\psi$ annihilation. This process  halts the further emission of GWs and, in the case of an IMD era, reestablishes the radiation era. Our approach, similar to Refs.~\cite{Dalianis:2020gup, Harada:2016mhb} will only focus on the nonlinear collapse process within the Zel'dovich approximation, and the gravitational signal which may exist due to the relaxation process will be left for future work.

Within the Zel'dovich formalism the coordinate of a particle is written as
\begin{equation}
r_i = a(t) q_i + b(t) p_i(q_i)
\end{equation}
where $a(t)$ is the usual scale factor encoding the expansion of the Universe, $q_i$ is the comoving coordinate, $b(t)$ is the growing mode describing the instability of the overdensity, and $p_i$ are deviation vectors that depend on the initial perturbation.

To study the motion of a group of particles around $q_i$, a deformation or strain tensor $D_{ik}$ is defined as
\begin{equation}
\begin{split}
D_{ik}
&=
\frac{\partial r_i}{\partial q_k} = a(t) \delta_{ik} + b(t) \frac{\partial p_i}{\partial q_k}\\[0.25cm]
&=
{\rm diag}(a - \alpha b, a - \beta b, a - \gamma b)
\end{split}
\end{equation}
where we have chosen a basis so that $\partial p_i/\partial q_k$ is diagonal,
\begin{equation}
\frac{\partial p_i}{\partial q_k}
=
-{\rm diag}(\alpha,\beta,\gamma)
.
\end{equation}
The mass contained within the Lagrangian volume is
\begin{equation}
\label{eq:MassEq}
M_\psi
=
\int \rho_\psi\ d^3 r
=
\bar{\rho}_\psi a^3 \int d^3 q
\end{equation}
where the difference between the two right-most expressions is the Jacobian determinant, i.e.,
\begin{equation}
\rho_\psi
(a - \alpha b)(a - \beta b)(a - \gamma b)
=
\bar{\rho}_\psi a^3
.
\end{equation}
We will also assume that the initial deviations are small, so that the perturbations are nearly spherical. Specifically, when a perturbation of size $q$ and wave number $k = q^{-1}$ enters the horizon at $t_q$,
\begin{equation}
\label{eq:HorzCross}
a(t_q)q = H^{-1}(t_q)
\end{equation}
we assume that $\alpha b(t_q)/a(t_q)\ll 1$, $\beta b(t_q)/a(t_q)\ll 1$, and $\gamma b(t_q)/a(t_q)\ll 1$ so that we can ensure that the entire ellipsoid is within the Hubble volume.
To leading order, the density contrast is given by~\cite{Dalianis:2020gup, Harada:2016mhb}
\begin{equation}
\label{eq:DensCont}
\delta_L \equiv
\left(
\frac{\rho_\psi - \bar{\rho}_\psi}{\bar{\rho}_\psi}
\right)_L
=
(\alpha + \beta + \gamma)\frac{b}{a}
.
\end{equation}
Note that for $b > 0$, we have $\delta_L > 0$ if and only if $\alpha + \beta + \gamma > 0$. As discussed previously, perturbations at most grow linearly with scale factor. The presence of a long-range scalar force enhances this growth and so, \textit{the key assumption of this work is that} $\delta_L\propto a^p$, $p\geq 1$. Crucially, the $p = 1$, $s = 2/3$ should reproduce precisely the work of Ref.~\cite{Dalianis:2020gup}, which we shall demonstrate later.

Under this assumption, Eq. \eqref{eq:DensCont} implies that $b\propto a^{p + 1}$ and therefore grows faster than the scale factor. The perturbation will collapse along at least one of the three axes. Without loss of generality we assume that
\begin{equation}
\label{eq:ParamConstI}
\alpha > 0,
\quad
-\infty < \gamma \leq \beta \leq \alpha < \infty,
\quad
\alpha + \beta + \gamma > 0
.
\end{equation}
Additionally, we can determine the timescale $t_q$ utilizing Eqs. \eqref{eq:MassEq} and \eqref{eq:HorzCross} to give
\begin{equation}
\label{eq:tq}
t_q = 2sGM_\psi
\left(
1 + \frac{\Omega_{\rm rad}(t_q)}{\Omega_\psi(t_q)}
\right)
\end{equation}
where $\Omega_X = \rho_X/\rho_{\rm crit}$. The horizon entry timescale is only one of three important events in our scenario. After entering the horizon, the perturbation grows until reaching some maximum size at $t_{\max}$. The overdensity subsequently collapses, at $t_{\rm col}$, due to the dominant scalar force. In this work, we will only examine the effects of scalar-induced collapse, though scalar cooling might dramatically decrease the time needed for the halo to collapse~\cite{Flores:2020drq, Flores:2021tmc}.

The derivation of $t_{\max}$ and $t_{\rm col}$ follows a similar methodology as in Ref.~\cite{Dalianis:2020gup} the details of which have been included in the Supplemental Material. We find that
\begin{equation}
t_{\max}
=
\left(
\frac{\alpha + \beta + \gamma}{(p + 1)\alpha\delta_L(t_q)}
\right)^{1/ps} t_q
\end{equation}
and
\begin{equation}
t_{\rm col} = (p + 1)^{1/ps} t_{\max}
.
\end{equation}

The deviations $\alpha$, $\beta$, $\gamma$ are determined by the Doroshkevich probability function as given by~\cite{Doroshkevich:1970spa},
\begin{equation}
\begin{split}
&\mathcal{F}_D(\alpha,\beta,\gamma,\sigma_3)
= -\frac{27}{8\sqrt{5}\pi\sigma_3^6}(\alpha-\beta)(\beta-\gamma)(\gamma-\alpha)\\
&\times \exp
\left[
- \frac{3}{5\sigma_3^2}
\left(
(\alpha^2
+ \beta^2 + \gamma^2)
-
\frac{1}{2}(\alpha\beta + \beta\gamma + \gamma\alpha)
\right)\right] 
\end{split}
\end{equation}
where $\alpha \geq \beta \geq \gamma$. The probability function is normalized to 1,
\begin{equation}
\int_{-\infty}^\infty
d\alpha
\int_{-\infty}^{\alpha}
d\beta
\int_{-\infty}^{\beta}
d\gamma\ 
\mathcal{F}_D(\alpha,\beta,\gamma,\sigma_3)
=
1
.
\end{equation}
We will denote the standard deviation of $\delta_L(t_q)$ as $\sigma$, given explicitly as
\begin{equation}
\sigma^2
=
\langle\delta_L^2(t_q)\rangle
=
\langle (\alpha + \beta + \gamma)^2 \rangle
\left.
\left(
\frac{b}{a}
\right)
\right|_{t = t_q} = 5\sigma_3^2
.
\end{equation}
For this expression, we fixed the normalization such that $b(t)/a(t) = [a(t)/a(t_q)]^p$. In addition to our restrictions Eq.~\eqref{eq:ParamConstI} we will also require that the maximum expansion time $t_{\rm max}$ is larger than the time of entry $t_q$. This defines a subspace $\mathcal{S}$,
\begin{equation}
\label{eq:IntSubspace}
\begin{split}
0 < \alpha &< \infty\\
-\frac{\alpha}{2}[1 - (p + 1)\sigma] < \beta &< \alpha\\
-\beta - \alpha [1 - (p + 1)\sigma] < \gamma &< \beta
\end{split}
\end{equation}
over which we will limit our calculations.

The stochastic GW background is traditionally characterized by the normalized energy density,
\begin{equation}
\Omega_{\rm GW}(t,f)
\equiv
\frac{1}{\rho_{\rm crit}}
\frac{d\rho_{\rm GW}}{d\ln f}
.
\end{equation}
Following a framework similar to Ref.~\cite{Dalianis:2020gup}, the present day GW energy density fraction reads as
\begin{equation}
\label{eq:MainOmGWEq}
\begin{split}
\Omega_{\rm GW}(t_0,f_0)
&=
\frac{1}{\rho_{\rm crit}(t_0)}
\frac{4\pi G}{5c^5}
\int_{\mathcal{S}}d^3{\boldsymbol \alpha}
\sum_{n = 1}^N \frac{1}{1 + z_n} 
\\[0.25cm]
&\times 
\sum_{ij}|\tilde{\dddot{Q}}_{ij}^n[2\pi f_0(1 + z_n)]|^2\\[0.25cm]
&\times
[2\pi f_0(1 + z_n)]
\left(\frac{4\pi}{3}q^3\right)^{-1}\\[0.25cm]
&\times
\mathcal{F}_D({\boldsymbol \alpha},\sigma)
\Theta[t_{\rm CO} - t_{\rm col}({\boldsymbol \alpha})].
\end{split}
\end{equation}
where the sum is over partitioned time intervals from the initial time to the final time of emission and $\tilde{\dddot{Q}}_{ij}$ is the Fourier transform of the third time derivative of the quadrupole moment. For notational simplicity, we have introduced the vector ${\boldsymbol \alpha} = (\alpha,\beta,\gamma)$. The Heaviside function enforces the condition that collapse occurs before some ``cut-off time." In the context of a RD era, this time could correspond to the lifetime of the constituent $\psi$ particles where in the IMD era $t_{\rm CO}$ is simply the time when radiation domination is reestablished.

In Eq. \eqref{eq:MainOmGWEq}, we have assumed that the mass distribution of overdensities is monochromatic for simplicity. The precise nature of the mass distribution resulting from long-range forces has yet to be determined, though it could possibly follow the Press-Schechter distribution which follows from gravitational forces~\cite{Press:1973iz, Flores:2020drq}.

As in Ref.~\cite{Dalianis:2020gup}, we will assume that all of the GWs are emitted in a single interval, i.e., $N = 1$. Partitioning the interval was motivated by the fact that GWs emitted in different subintervals will be redshifted different amounts as they propagate. For simplicity, we assume that the GWs are emitted instantaneously at $t_{\rm col}$. In view of the fact that $t_{\rm col} = (p + 1)^{1/ps}t_{\max}$, with $p \geq 1$, we see that this approximation not only is reasonable, but becomes better as $p$ increases. In this approximation the final expression for $\Omega_{\rm GW}(t_0, f_0)$ is
\begin{equation}
\label{eq:FinalOmGWEq}
\begin{split}
&\Omega_{\rm GW}(t_0,f_0)
=
\frac{2\pi f_0}{\rho_{\rm crit}(t_0)}\left(\frac{4\pi}{3}q^3\right)^{-1}
\frac{4\pi G}{5c^5}\\[0.25cm]
&\times
\int_{\mathcal{S}}d^3{\boldsymbol \alpha}
\sum_{ij}|\tilde{\dddot{Q}}_{ij}\{2\pi f_0[1 + z_{\rm col}({\boldsymbol\alpha })]\}|^2\\[0.25cm]
&\times
\{2\pi f_0[1 + z_{\rm col}({\boldsymbol\alpha })]\}
\mathcal{F}_D({\boldsymbol \alpha},\sigma)
\Theta[t_{\rm CO} - t_{\rm col}({\boldsymbol \alpha})].
\end{split}
\end{equation}
Here, the redshift $z_{\rm col}$ is related to the collision timescale through
\begin{equation}
1 + z_{\rm col} = \frac{1 + z_{\rm CO}}{t_{\rm col}^{s}}
\left(
\frac{3s^2}{8\pi G\rho_{\rm CO}}
\right)^{s/2}
\end{equation}
where $z_{\rm CO}$ is the redshift at $T_{\rm CO}$ and $\rho_{\rm CO} = \pi^2 g_\star T_{\rm CO}^4/30$. We may also use these quantities to express the $q$, the comoving radius of the perturbation at the time of horizon entry, $t_q$~\cite{Dalianis:2020gup},
\begin{equation}
\begin{split}
&q^{-1}(M,T_{\rm CO})
=
\left(
\frac{3M(1 + z_{\rm CO})^3}{4\pi \rho_{\rm CO}}
\right)^{-1/3}\\
&\simeq
1.2\times 10^{10}\ \text{Mpc}^{-1}
\left(
\frac{T_{\rm CO}}{10^{10}\ {\rm GeV}}
\right)^{\frac{1}{3}}
\left(
\frac{M}{M_\odot}
\right)^{-\frac{1}{3}}
\end{split}
\end{equation}
where $z_{\rm CO}$ was determined using conservation of entropy. Our selection of $T_{\rm CO}$ will be informed by the Heaviside function in Eq.~\eqref{eq:FinalOmGWEq}. In particular,
\begin{equation}
\begin{split}
T_{\rm CO}
&\lesssim
\frac{0.2\ {\rm GeV}}{\left(
1 + \frac{\Omega_{\rm rad}(t_q)}{\Omega_\psi(t_q)}
\right)^{1/2}}
\left(
\frac{M_\psi}{M_\odot}
\right)^{-1/2}
\left(
\frac{g_*}{106.75}
\right)^{-1/4}\times \\
&\quad 
\times 
\left(
\frac{\alpha\sigma}{\alpha + \beta + \gamma}
\right)^{1/(2ps)}.
\end{split}
\end{equation}
In the RD case, $\Omega_{\rm rad}\gg \Omega_{\psi}$ greatly restricts possible values of $T_{\rm CO}$.

Fundamental to Eq.~\eqref{eq:FinalOmGWEq} is the mass quadruple moment. We have included a full derivation of $Q_{ij}$ in the Supplemental Material. For simplicity, we will simply present the Fourier transform:
\begin{multline}
\label{eq:GenQuad}
\tilde{\dddot{Q}}_{ii}(\omega)
=
\frac{1}{2\pi}\int_{t_1}^{t_2} \dddot{Q}_{ii}(t) e^{-i\omega t}\ d t\\[0.25cm]
=\frac{2M}{30\pi(p + 1)s} \frac{t_q^{2(1 - s)}}{t_{\max}^{2ps}}
\left[
\mathcal{A}_{ps} v_i({\boldsymbol \alpha})\ t_{\max}^{ps}\ \mathcal{I}_1(t_1,t_2,\omega)
\right.\\
\left.
-
\mathcal{B}_{ps} w_i({\boldsymbol \alpha})\ \mathcal{I}_2(t_1,t_2,\omega)
\right]
\end{multline}
where
\begin{equation}
\label{eq:IntDefs}
\begin{split}
\mathcal{I}_1(t_1,t_2,\omega)
&=
\int_{t_1}^{t_2}
t^{(p + 2)s - 3}
e^{-i\omega t}\ d t,\\[0.25cm]
\mathcal{I}_2(t_1,t_2,\omega)
&=
\int_{t_1}^{t_2}
t^{2(p + 1)s - 3}
e^{-i\omega t}\ d t.
\end{split}
\end{equation}
and the expressions for the constants  $\mathcal{A}_{ps}$, $\mathcal{B}_{ps}$ and the vectors ${\bf v}({\boldsymbol \alpha})$ and ${\bf w}({\boldsymbol \alpha})$ are defined alongside their derivation in the Supplemental Material. The square of Eq.~\eqref{eq:GenQuad} with $(t_1,t_2)\to (t_{\max},t_{\rm col})$, and $\omega\to 2\pi f_0(1 + z_{\rm col})$, completes our derivation of $\Omega_{\rm GW}(t_0,f_0)$.

\begin{figure}[htb]
\centering
\includegraphics[width=0.95\linewidth]{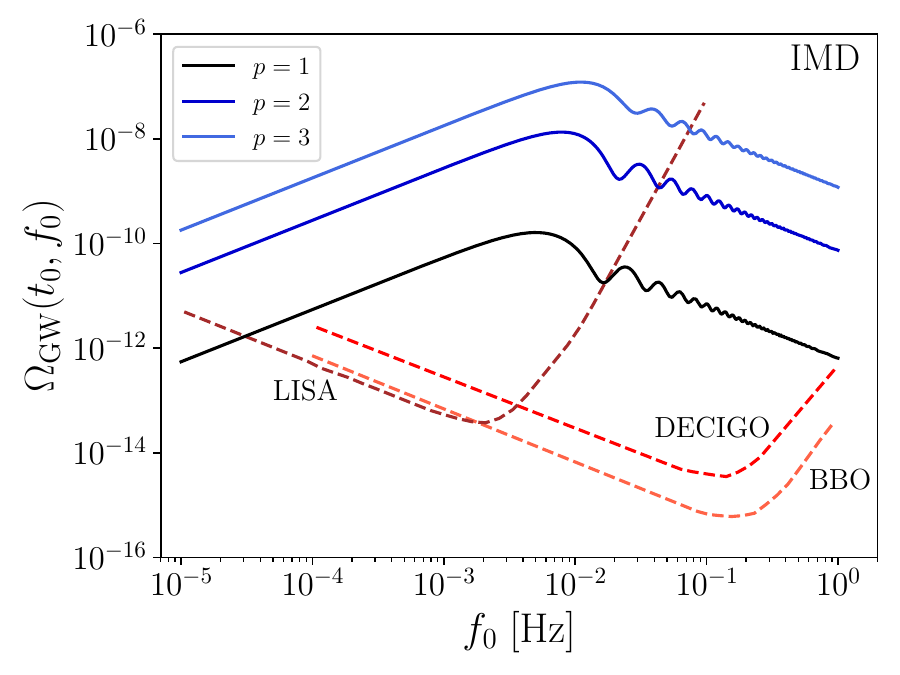}
\caption{The GW signal for varying $p$ in an intermediate matter dominated era. Here, we take $T_{\rm CO} = 10^4$ GeV, $\sigma = 10^{-1}$. Additionally, we  assume a monochromatic mass function such that all the overdensities have mass $M_\psi = 10^{-12}\ M_\odot$. Sensitivity curves for LISA~\cite{LISA:2017pwj}, DECIGO~\cite{Seto:2001qf, Sato:2017dkf}, and BBO~\cite{Corbin:2005ny} are shown.}
\label{fig:IMDPlot}
\end{figure}

Estimating $\Omega_{\rm GW}(t_0,f_0)$ requires numerical integration over $(\alpha,\beta,\gamma)$ subject to the restrictions specified in Eq.~\eqref{eq:IntSubspace}. To begin, Fig.~\ref{fig:IMDPlot} illustrates the enhanced amplitude resulting from varying values of $p$ in an IMD era. The enhancement of the peak signal $\Omega_{\rm GW}^{\rm peak}\equiv \Omega_{\rm GW}(t_0, f_{\rm peak})$ is quite dramatic, even for modest values of $p > 1$, with $p = 1$ being consistent with the result obtained in Ref.~\cite{Dalianis:2020gup}. We note that the sizable enhancement may conflict with indirect bounds derived for the integrated quantity, $\Omega_{\rm GW}\equiv \int \Omega_{\rm GW}(f)\ d \ln f$~\cite{Lasky:2015lej, LIGOScientific:2016jlg, Caldwell:2022qsj}. 

\begin{figure*}[htb]
\centering
\includegraphics[width=0.95\linewidth]{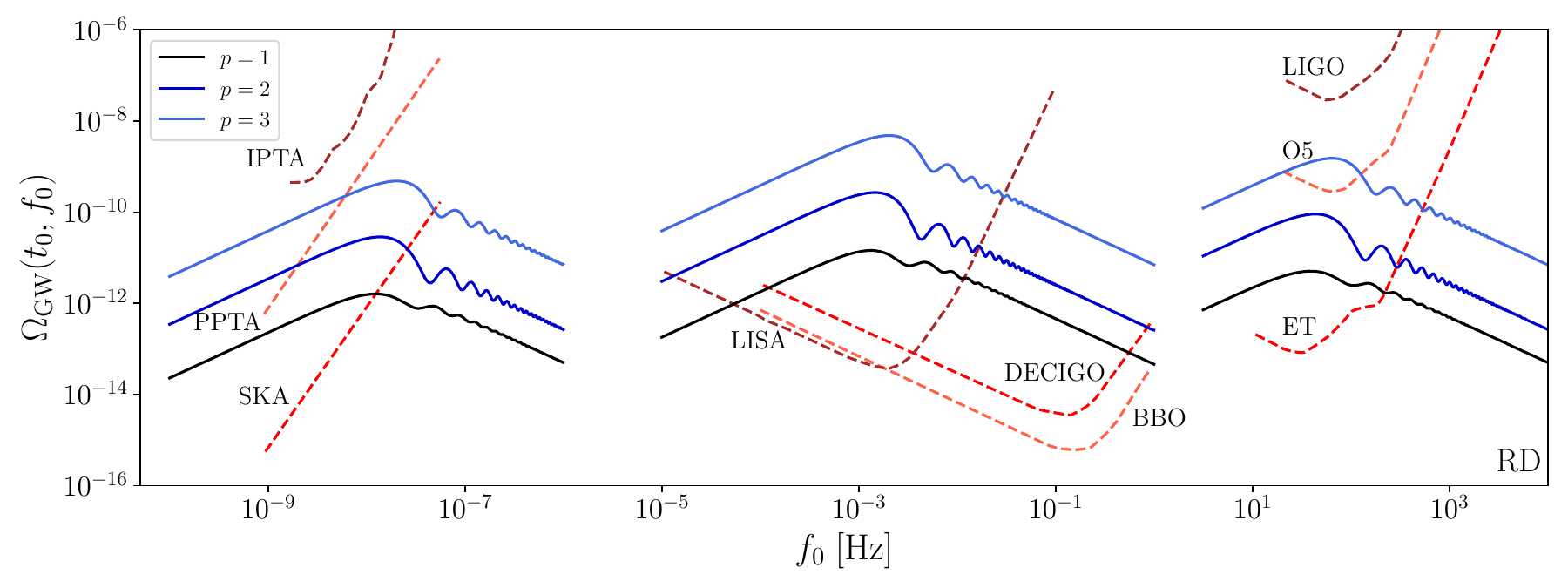}
\caption{The GW signal across a broad range of frequencies with varying $p$, during a radiation dominated era. In all cases $\sigma = 10^{-1}$ and $\Omega_{\rm rad}/\Omega_\psi  = 100$. Again, we  assume a monochromatic mass distribution. The masses and cut-off temperatures are (left) $10^{-2}\ M_\odot$ and $10^{-2}$ GeV, (center) $10^{-12}\ M_\odot$ and $10^{4}$ GeV, (right) $10^{-21}\ M_\odot$ and $10^9$ GeV, respectively. Sensitivity curves for PTA observations~\cite{Thorsett:1996dr, Saito:2008jc, Bugaev:2010bb, Chen:2019xse}, SKA~\cite{Janssen:2014dka, Maartens:2015mra} LIGO~\cite{TheLIGOScientific:2016dpb}, ET~\cite{Sathyaprakash:2012jk}, LISA~\cite{LISA:2017pwj}, DECIGO~\cite{Seto:2001qf, Sato:2017dkf}, and BBO~\cite{Corbin:2005ny}  are shown.}
\label{fig:RDPlot}
\end{figure*}
Given the complexity of $\Omega_{\rm GW}(t_0, f_0)$, we can determine an approximate parametrization of $\Omega_{\rm GW}^{\rm peak}$ and $f_{\rm peak}$ after computing the full signal for various combinations of parameters. We select a parametrization of the form,
\begin{equation}
\begin{split}
\Omega_{\rm GW}^{\rm peak}
&\approx
\Omega_{\rm GW,0}^{\rm peak}\ \sigma^{\delta_A/p}
\left(
\frac{p}{3}
\right)^{\alpha_A}
\left(
\frac{T_{\rm CO}}{10^4\ {\rm GeV}}
\right)^{\beta_A}
\times
\\[0.25cm]
&\times
\left(
\frac{M}{10^{-12}\ M_\odot}
\right)^{\gamma_A}
\left(
\frac{\Omega_{\rm rad}(t_q)/\Omega_{\psi}(t_q)}{100}
\right)^{\varepsilon_A}
\end{split}
\end{equation}
and
\begin{equation}
\begin{split}
f_{\rm peak}
&\approx
f_{\rm peak, 0}\ 
\left(
\frac{p}{3}
\right)^{\alpha_f}
\left(
\frac{T_{\rm CO}}{10^4\ {\rm GeV}}
\right)^{\beta_f}\\[0.25cm]
&\times
\left(
\frac{M}{10^{-12}\ M_\odot}
\right)^{\gamma_f}
\left(
\frac{\Omega_{\rm rad}(t_q)/\Omega_{\psi}(t_q)}{100}
\right)^{\varepsilon_f}
\end{split}
\end{equation}
where the coefficients for each formula are presented in Tables~\ref{tab:AmplCoeff} and \ref{tab:FreqCoeff}, respectively.

\begin{table}[htb!]
\begin{tabular}{c|c|c|c|c|c|c}
\hline
   & $\Omega_{\rm GW,0}^{\rm peak}$ &   $\alpha_A$	&  $\beta_A$ & $\gamma_A$ & $\delta_A$ & $\varepsilon_A$\\\hline
MD &     $1.4\times 10^{-7}$       &       5       &    $4/3$   &	  $2/3$   &     1      & 0              \\\hline
RD &     $4.4\times 10^{-9}$       & 	   7       &    $1$     &	  $1/2$   &    $p/5$   & $-1/2$         \\\hline
\end{tabular}
\caption{Coefficients for parametrization of $\Omega_{\rm GW}^{\rm peak}$.}
\label{tab:AmplCoeff}
\end{table}
\begin{table}[htb!]
\begin{tabular}{c|c|c|c|c|c}
\hline
   & $f_{\rm peak,0}\ [{\rm Hz}]$ & $\alpha_f$ &  $\beta_f$ & $\gamma_f$  & $\varepsilon_f$ \\\hline
MD &      $1.1\times 10^{-2}$    &     5/6    &    $1/3$   &   $-1/3$    &        0        \\\hline
RD &      $2.0\times 10^{-3}$    & 	   2/5    &    $0$     &   $-1/2$    &      $-1/2$     \\\hline
\end{tabular}
\caption{Coefficients for parametrization of peak frequency $f_{\rm peak}$.}
\label{tab:FreqCoeff}
\end{table}

In either era, the peak frequency at emission corresponds roughly to $f_{\rm peak}^{\rm emit} \sim 1/t_{\max}$. For a RD era, illustrated in Fig.~\ref{fig:RDPlot}, there is a notable suppression in amplitude compared with the IMD case. We also note in this new context that $\delta\propto a^p$, $p \geq 1$ always exceeds the logarithmic growth enabled by gravity only during a radiation era. 

The uniqueness of the long-range scalar forces in this scenario can be deduced in many ways. If the evolution of the Universe is known to have been RD from reheating until big bang nucleosynthesis, the presence of a signal of the type featured in Fig.~\ref{fig:RDPlot} can only be the result of the presence of an additional long-range scalar force. Without knowledge of the precise evolution of the Universe, we note that the infrared oscillations for the RD are suppressed as compared with the IMD case. The damping of these oscillations only occurs for variations in $p$ which helps reduce the degeneracy within our set of parameters.

The GW spectrum calculated scales as
\begin{equation}
\Omega_{\rm GW}(t_0, f_0)\propto
\begin{cases}
f_0 &\text{for}\ f_0\ll 1\\
f_0^{-1} &\text{for}\ f_0\gg 1\\
\end{cases}
\end{equation}
as also demonstrated in Ref.~\cite{Dalianis:2020gup}; however, there is reason to potentially doubt this scaling far from the peak amplitude. By examining a finite time interval, $[t_{\max}, t_{\rm col}]$, we essentially are windowing the quadrupole moment with a step function. The net result leads to a Fourier transform proportional to a constant in the IR, and scaling as $1/\omega^{2}$ in the UV regardless of the functional form of $Q_{ij}(t)$. Furthermore, this result is inconsistent with the expectation that the GW amplitude should scale as $\omega^3$ for small frequencies during emission during a RD era~\cite{Cai:2019cdl}. While the solution near the peak is reliable, in future, it would be advantageous to investigate scaling behavior through alternative methods including numerical simulations and cosmological perturbation theory.

In this work, we investigated the gravitational waves produced from the evolution of nonspherical density perturbations in the presence of a long-range scalar force. We consider the scenario where heavy, nonrelativistic fermions interact via an attractive Yukawa interaction. This force, which can be significantly stronger than gravity, causes structure to form rapidly. The subsequent collapse of these aspherical overdensities, which could lead to PBHs, has a nonvanishing quadrupole moment allowing for the emission of GWs. This signal, which fits inside the sensitivity curves of many future GW observatories, can be significantly stronger than the gravity-only scenario and would be distinguishable from other signals which occur during a RD era~\cite{Mollerach:2003nq, Ananda:2006af, Baumann:2007zm, DeLuca:2019llr}.  If the halos form but do not collapse into black holes (due to long cooling timescales), the resulting density perturbations can help seed structure formation in a matter dominated era.  For the parameters we discuss, the effect would be limited to extremely small length scales, but some generalizations of our mechanism can have implications for large-scale structure formation.

\begin{acknowledgments}
We thank I.~Dalianis, G.~Dom\`enech, and D.~Inman for helpful discussions.  
This work  was  supported  by the U.S. Department of Energy (DOE) Grant No.  DE-SC0009937.   The work of A.K.  was  also supported  by  the World Premier International Research Center Initiative (WPI),  MEXT,  Japan, and by Japan Society for the Promotion of Science (JSPS) KAKENHI Grant No. JP20H05853. M.S. is supported by the JSPS KAKENHI Grant No. 19H01895, No. 20H04727, and No. 20H05853. This work used computational and storage services associated with the Hoffman2 Shared Cluster provided by UCLA Institute for Digital Research and Education’s Research Technology Group.

\end{acknowledgments}


\bibliography{biblio}


\appendix

\section{Supplemental Material}

\subsection{Derivation of time scales}

We begin by first deriving $t_{\max}$. First, recall the expression for the density contrast to linear order in the deviations $\{\alpha,\beta,\gamma\}$,
\begin{equation}
	\delta_L \equiv
	\left(
	\frac{\rho_\psi - \bar{\rho}_\psi}{\bar{\rho}_\psi}
	\right)_L
	=
	(\alpha + \beta + \gamma)\frac{b}{a}
	.
\end{equation}
We will utilize the continuity of $\alpha$, $\beta$ and $\gamma$ so that we can locally take the coordinates $q_i$ to be~\cite{Dalianis:2020gup, Harada:2016mhb}
\begin{equation}
	\label{eq:ModCoords}
	\begin{cases}
		r_1 = (a - \alpha b)q_1\\
		r_2 = (a - \beta b)q_2\\
		r_3 = (a - \gamma b)q_3.
	\end{cases}
\end{equation}
By definition, $t_{\max}$ is the time that satisfies
\begin{equation}
	\dot{r}_{1}(t_{\rm max}) = 0
\end{equation}
which implies $\dot{b}(t_{\rm max})/\dot{a}(t_{\rm max})  = 1/\alpha$ and therefore,
\begin{equation}
	\label{eq:tmaxRatio}
	\frac{b(t_{\max})}{a(t_{\max})} = \frac{1}{(p + 1)\alpha}
	.
\end{equation}
Combining \eqref{eq:DensCont} evaluated at $t_q$ and \eqref{eq:tmaxRatio}, we find that
\begin{equation}
	t_{\max}
	=
	\left(
	\frac{\alpha + \beta + \gamma}{(p + 1)\alpha\delta_L(t_q)}
	\right)^{1/ps} t_q.
\end{equation}
Next, we focus our attention on $t_{\rm col}$. Collapse takes place when $r_1(t_{\rm col}) = 0$. This naturally implies $b(t_{\rm col})/a(t_{\rm col}) = 1/\alpha$. Using this result in combination with \eqref{eq:tmaxRatio} implies
\begin{equation}
	t_{\rm col} = (p + 1)^{1/ps} t_{\max}
	.
\end{equation}

\subsection{Derivation of the quadrupole moment}

We can express $Q_{ij}$ in terms of the moment of inertia,
\begin{equation}
	\label{eq:QuadMom}
	Q_{ij}(t)
	=
	-
	I_{ij}(t)
	+
	\frac{1}{3}\delta_{ij}{\rm Tr} I(t)
	.
\end{equation}
In the reference frame of the principle axes, the moment of inertia is
\begin{equation}
	\label{eq:MomIner}
	I_{ij}
	=
	\frac{M}{5}
	\begin{pmatrix}
		r_2^2 + r_3^2 & 0 & 0\\
		0 & r_1^2 + r_3^2 & 0\\
		0 & 0 & r_1^2 + r_2^2\\
	\end{pmatrix}
	.
\end{equation}
Using \eqref{eq:ModCoords} and \eqref{eq:tmaxRatio} we find,
\begin{equation}
	\label{eq:RadDirs}
	\begin{split}
		r_1(t) &= \frac{1}{s}t_q^{1 - s}t^s
		\left(
		1 - \frac{1}{(p + 1)}
		\left(
		\frac{t}{t_{\max}}
		\right)^{ps}
		\right),\\[0.25cm]
		r_2(t) &= \frac{1}{s}t_q^{1 - s}t^s
		\left(
		1 - \frac{\beta}{(p + 1)\alpha}
		\left(
		\frac{t}{t_{\max}}
		\right)^{ps}
		\right),\\[0.25cm]
		r_3(t) &= \frac{1}{s}t_q^{1 - s}t^s
		\left(
		1 - \frac{\gamma}{(p + 1)\alpha}
		\left(
		\frac{t}{t_{\max}}
		\right)^{ps}
		\right).
	\end{split}
\end{equation}
Combining \eqref{eq:QuadMom} - \eqref{eq:RadDirs} we obtain
\begin{equation}
	\begin{split}
		\dddot{Q}_{ii}(t)
		=
		&\frac{2M}{15(p + 1)s}
		\frac{t_q^{2(1 - s)}}{t_{\max}^{2ps}}t^{(p + 2)s - 3}\\[0.25cm]
		&\hspace{1cm}
		\times 
		\left[
		\mathcal{A}_{ps}v_i({\boldsymbol \alpha})t_{\max}^{ps}
		- 
		\mathcal{B}_{ps}w_i({\boldsymbol \alpha})t^{ps}
		\right].
	\end{split}
\end{equation}
The constants $\mathcal{A}_{ps}$ and $\mathcal{B}_{ps}$ utilized in \eqref{eq:GenQuad} are
\begin{equation}
	\begin{split}
		\mathcal{A}_{ps}
		&=
		(p + 2)(s(p + 2) - 2)(s(p + 2) - 1)\\[0.25cm]
		\mathcal{B}_{ps}
		&=
		2(ps + s - 1)(2s(p + 1) - 1).
	\end{split}
\end{equation}
we note that for $p = 1$ and $s = 2/3$, $\mathcal{A}_{ps} = 0$, thus simplifying much of the expressions above. The vector expressions ${\bf v}({\boldsymbol \alpha})$ and ${\bf w}({\boldsymbol \alpha})$ are defined as,
\begin{widetext}
	\begin{equation}
		\begin{split}
			{\bf v}({\boldsymbol \alpha})
			&=
			\left\langle
			-2 + \frac{\beta}{\alpha} + \frac{\gamma}{\alpha},\ 
			1 - \frac{2\beta}{\alpha} + \frac{\gamma}{\alpha},\ 
			1 + \frac{\beta}{\alpha} - \frac{2\gamma}{\alpha}
			\right\rangle
			,\\[0.25cm]
			{\bf w}({\boldsymbol \alpha})
			&=
			\left\langle
			-2 + \left(\frac{\beta}{\alpha}\right)^2 + \left(\frac{\gamma}{\alpha}\right)^2,\ 
			1 - 2\left(\frac{\beta}{\alpha}\right)^2 + \left(\frac{\gamma}{\alpha}\right)^2,
			1 + \left(\frac{\beta}{\alpha}\right)^2  - 2\left(\frac{\gamma}{\alpha}\right)^2.
			\right\rangle
		\end{split}
	\end{equation}
	The Fourier transform of the above quadrupole moment is given as,
	\begin{align}
		\tilde{\dddot{Q}}_{ii}(\omega)
		&=
		\frac{1}{2\pi}\int_{t_1}^{t_2} \dddot{Q}_{ii}(t) e^{-i\omega t}\ d t\\[0.25cm]
		&=
		\frac{2M}{30\pi(p + 1)s} \frac{t_q^{2(1 - s)}}{t_{\max}^{2ps}}
		\left[
		\mathcal{A}_{ps} v_i({\boldsymbol \alpha})\ t_{\max}^{ps}\ \mathcal{I}_1(t_1,t_2,\omega)
		-
		\mathcal{B}_{ps} w_i({\boldsymbol \alpha})\ \mathcal{I}_2(t_1,t_2,\omega)
		\right]
	\end{align}
	where
	\begin{equation}
		\begin{split}
			\mathcal{I}_1(t_1,t_2,\omega)
			&=
			\int_{t_1}^{t_2}
			t^{(p + 2)s - 3}
			e^{-i\omega t}\ d t,\\[0.25cm]
			\mathcal{I}_2(t_1,t_2,\omega)
			&=
			\int_{t_1}^{t_2}
			t^{2(p + 1)s - 3}
			e^{-i\omega t}\ d t.
		\end{split}
	\end{equation}
	Lastly, we may also rewrite the integral expressions \eqref{eq:IntDefs} in terms of the the generalized exponential integral
	\begin{equation}
		E_n(x) = \int_1^\infty \frac{e^{-xt}}{t^n}\ dt
		.
	\end{equation}
	Using this definition integrals \eqref{eq:IntDefs} read,
	\begin{equation}
		\begin{split}
			\mathcal{I}_1(t_1, t_2,\omega)
			&=
			\frac{1}{t_1^{2-(p + 2)s}}
			\left[
			E_{3-(p + 2)s}(i\omega t_1)
			-
			\left(\frac{t_1}{t_2}\right)^{2-(p + 2)s}
			E_{3-(p + 2)s}(i\omega t_2)
			\right]\\[0.25cm]
			\mathcal{I}_2(t_1, t_2,\omega)
			&=
			\frac{1}{t_1^{2-2(p + 1)s}}
			\left[
			E_{3-2(p + 1)s}(i\omega t_1)
			-
			\left(\frac{t_1}{t_2}\right)^{2-2(p + 1)s}
			E_{3-2(p + 1)s}(i\omega t_2)
			\right].
		\end{split}
	\end{equation}

\end{widetext}

\subsection{Derivation of GW Spectrum}

Here we estimate the GW signal produced in the epoch of interest due to the collapse of a density perturbation with wavelength $q$ that enters the horizon at time $t_q$. This derivation follows Ref.~\cite{Dalianis:2020gup} closely. 

At the time of entry $t_q$, the Hubble volume is given by $V_q = (4/3)\pi q^3$. The current comoving volume is $V_{\rm com}(t_0) = (4/3)\pi H^{-3}(t_0)$. The number of volumes of size $V_q$ is $H^{-3}/q^3$. The differential number of sources with deformation parameters $(\alpha, \alpha + d \alpha)$, $(\beta, \beta + d \beta)$ and $(\gamma, \gamma + d \gamma)$ is
\begin{equation}
	d N
	=
	\frac{V_{\rm com}(t)}{\frac{4}{3}\pi q^3}\mathcal{F}_D(\alpha,\beta,\gamma,\sigma)\ d \alpha\ d\beta\ d \gamma
	.
\end{equation}
The power emitted in the form of GWs from each Hubble patch that encloses a perturbation with a quadrupole tensor $Q_{ij}$ is
\begin{equation}
	\frac{d E_e}{d t}
	=
	\frac{G}{5c^5}
	\sum_{ij}
	\dddot{Q}_{ij}(t)
	\dddot{Q}_{ji}(t)
	.
\end{equation}
Let us define an interval between $t_i$ and $t_f$ which encapsulates the time interval where GW are emitted. Consider we separate the interval $[t_i, t_f]$ into $N$ intervals of size $\delta t$. For each interval, we expand the quadrupole in Fourier series,
\begin{equation}
	Q_{ij}(t)
	=
	\sum_{n = -\infty}^\infty
	\tilde{Q}_{ij}^N(n\omega_0) e^{-in\omega_0 t}
\end{equation}
where $\omega_0 = 2\pi/\delta t$ and $t\in [t_N, t_N + \delta t]$. The $\tilde{Q}$ components are
\begin{equation}
	\tilde{Q}_{ij}^N(n\omega_0)
	=
	\frac{\omega_0}{2\pi}
	\int_{t_N}^{t_N + \delta t}
	Q_{ij}(t) e^{-in\omega_0 t}\ d t
	.
\end{equation}
We can take the continuum limit for $\omega$, we transform the last two equations to
\begin{equation}
	\begin{split}
		Q_{ij}
		&=
		\int \tilde{Q}_{ij}^N e^{i\omega t}\ d \omega,\\[0.25cm]
		\tilde{Q}_{ij}^N(\omega)
		&=
		\frac{1}{2\pi}\int_{t_N}^{t_N + \delta t}
		Q_{ij}(t) e^{-i\omega t}\ d t
	\end{split}
\end{equation}
\begin{widetext}
	The emitted energy $E_e^N$ in the bin $[t_{N_i}, t_{N_i} + \delta t]$ is

	\begin{equation}
		\begin{split}
			E_e^N
			&=
			\frac{2G}{5c^5}
			\int_{t_N}^{t_N + \delta t}
			\sum_{ij}\sum_{nm}
			(in\omega)^3 (-im\omega)^3
			\times
			\tilde{Q}_{ij}^N(n\omega)\tilde{Q}_{ij}^{*N}(m\omega)e^{-i(n - m)\omega t}\ d t\\[0.25cm]
			&=
			\frac{2G}{5c^5}
			\sum_{ij}\sum_n
			(n\omega)^6 
			\tilde{Q}_{ij}^N(n\omega)\tilde{Q}_{ij}^{*N}(n\omega)\delta t
		\end{split}
	\end{equation}
	The factor of $2$ is due to the restriction to positive $n$ and $m$ since the negative factors contribute equally. In the continuum limit,
	\begin{equation}
		d E_e^N
		=
		\frac{4\pi G}{5c^5}\omega^6
		\sum_{ij}
		\left|\tilde{Q}_{ij}^N\right|^2\ d \omega
	\end{equation}
	so that,
	\begin{equation}
		\frac{d E_e^N}{d \ln \omega}
		=
		\frac{4\pi G}{5c^5}\omega^7
		\sum_{ij}
		\left|\tilde{Q}_{ij}^N(\omega)\right|^2
	\end{equation}
	within the time bin $[t_N, t_N + \delta t]$. The differential energy is then given by,
	\begin{equation}
		d E_{\rm GW}({\boldsymbol \alpha})
		=
		\sum_N
		\frac{1}{1 + z_N}
		\frac{4\pi G}{5c^5}\omega^7
		\sum_{ij}
		\left|\tilde{Q}_{ij}^N(\omega)\right|^2
		\frac{V_{\rm com}(t_0)}{\frac{4}{3}\pi q^3}\mathcal{F}_D({\boldsymbol \alpha},\sigma)\ d^3{\boldsymbol \alpha}\ d \ln\omega 
	\end{equation}
	where we have defined ${\boldsymbol \alpha} = (\alpha,\beta,\gamma)$ for simplicity. Here, the energies emitted in each time interval is redshifted individually. Furthermore, the number of sources is approximately given by
	\begin{equation}
		N_{\rm tot}
		\simeq
		\int
		\frac{V_{\rm com}(t_0)}{\frac{4}{3}\pi q^3}
		\Theta(t_{\rm rh} - t_{\rm col}({\boldsymbol \alpha}))
		\mathcal{F}_D({\boldsymbol \alpha},\sigma)\ 
		d^3{\boldsymbol \alpha}
		.
	\end{equation}
	The differential energy per observed logarithmic frequency interval and deformation configuration is
	\begin{equation}
		\begin{split}
			\frac{
				d \rho_{\rm GW}(t_0, f_0)
			}{
				d^3{\boldsymbol \alpha}\ d\ln f_0
			}
			&=
			\sum_N
			\frac{1}{1 + z_N}
			\frac{4\pi G}{5c^5}
			\sum_{ij}
			\left|
			\tilde{Q}_{ij}^N (2\pi f_0(1 + z_N))
			\right|^2\\[0.25cm]
			&\times
			\left(2\pi f_0(1 + z_N)\right)^7\Theta\left(t_{\rm rh} - t_{\rm col}({\boldsymbol \alpha})\right)
			\left(
			\frac{4\pi}{3}q^{3}
			\right)^{-1}
			\mathcal{F}_D({\boldsymbol \alpha},\sigma_3)
		\end{split}
	\end{equation}
	In terms of the usual density parameter,
	\begin{equation}
		\Omega_{\rm GW}(t,f) 
		\equiv
		\frac{1}{\rho_{\rm crit}}
		\frac{d \rho_{\rm GW}}{d \ln f}
		.
	\end{equation}
	The present time value is then
	\begin{multline}
		\label{eq:FullGWEqn}
		\Omega_{\rm GW}(t_0,f_0)
		=
		\frac{1}{\rho_{\rm crit,0}}
		\int_{\mathcal{S}}
		d^3{\boldsymbol \alpha}
		\sum_N
		\frac{1}{1 + z_N}
		\frac{4\pi G}{5c^5}
		\sum_{ij}
		\left|
		\tilde{Q}_{ij}^N (2\pi f_0(1 + z_N))
		\right|^2\\
		\times
		\left(
		2\pi f_0(1 + z_N)
		\right)^7
		\Theta\left(t_{\rm rh} - t_{\rm col}({\boldsymbol \alpha})\right)
		\left(
		\frac{4\pi}{3}q^{3}
		\right)^{-1}
		\mathcal{F}_D({\boldsymbol \alpha},\sigma_3)
		.
	\end{multline}
	This is the \textit{key result} of Ref. \cite{Dalianis:2020gup}. Making use of the fact that
	\begin{equation}
		\tilde{\dddot{Q}}_{ij}(\omega)
		\sim
		\omega^3 \tilde{Q}_{ij}
	\end{equation}
	we recover the result as presented in the main text.

\end{widetext}

\end{document}